\documentclass[12pt]{iopart}
\usepackage[utf8]{inputenc}
\usepackage{graphicx}
\usepackage{multirow}
\usepackage{url}
\usepackage{iopams}
\usepackage{cite}
\usepackage{color}

\newcommand{\figurewidth}{0.84\textwidth}

\DeclareSymbolFont{extraup}{U}{zavm}{m}{n}
\DeclareMathSymbol{\vardiamond}{\mathalpha}{extraup}{87}
\begin{document}

\title[Photorecombination of berylliumlike  and boronlike silicon ions]{Absolute rate coefficients for photorecombination of berylliumlike and boronlike silicon ions}
\author{D.~Bernhardt$^1$,
A.~Becker$^2$,
C.~Brandau$^{3,4}$,
M.~Grieser$^2$,
M.~Hahn$^5$,
C.~Krantz$^2$,
M.~Lestinsky$^4$,
O.~Novotn\'y$^{2,5}$,
R.~Repnow$^2$,
D.~W.~Savin$^5$,
K.~Spruck$^{1,2}$,
A.~Wolf$^{\,2}$,
A.~M\"uller$^1$,
S.~Schippers$^{3}$}

\address{$^1$ Institut f\"ur Atom- und Molek\"ulphysik, Justus-Liebig-Universit\"at Gie{\ss}en, Leihgesterner Weg 217, 35392 Giessen, Germany}
\address{$^2$ Max-Planck-Institut f\"ur Kernphysik, Saupfercheckweg 1, 69117 Heidelberg, Germany}
\address{$^3$ I. Physikalisches Institut, Justus-Liebig-Universit{\"a}t Gie{\ss}en, Heinrich-Buff-Ring 16, 35392 Giessen, Germany}
\address{$^4$ GSI Helmholtzzentrum f\"ur Schwerionenforschung, Planckstra{\ss}e 1, 64291 Darmstadt, Germany}
\address{$^5$ Columbia Astrophysics Laboratory, Columbia University, New York, NY 10027, USA}
\ead{Stefan.Schippers@physik.uni-giessen.de}

\begin{abstract}
We report measured rate coefficients for electron-ion recombination for Si$^{10+}$ forming Si$^{9+}$ and for Si$^{9+}$ forming Si$^{8+}$, respectively.
The measurements were performed using the electron-ion merged-beams technique at a heavy-ion storage ring. Electron-ion collision energies ranged from 0 to 50~eV for Si$^{9+}$ and from 0 to 2000~eV for Si$^{10+}$, thus, extending previous measurements for Si$^{10+}$ [Orban \etal 2010 \textit{Astrophys.~J.} \textbf{721} 1603] to much higher energies. Experimentally derived rate coefficients for the recombination of Si$^{9+}$ and Si$^{10+}$ ions in a plasma are presented along with simple parameterizations. These rate coefficients are useful for the modeling of the charge balance of silicon in photoionized plasmas (Si$^{9+}$ and Si$^{10+}$) and in collisionally ionized plasmas (Si$^{10+}$ only). In the corresponding temperature ranges, the experimentally derived rate coefficients agree with the latest corresponding theoretical results within the experimental uncertainties.
\end{abstract}

\maketitle

\section{Introduction} \label{sec:intro}

In the universe, silicon is produced by stellar nucleosynthesis and in supernova explosions. It is the eighth most abundant element in the solar system \cite{Asplund2009} and also contributes significantly to the elemental composition of the intergalactic medium \cite{Aguirre2004a}. Consequently, prominent photoemission lines from silicon ions are observed from many cosmic plasmas such as the solar corona \cite{Doschek2010}, stellar atmospheres \cite{Guedel2009}, supernova remnants \cite{Decourchelle2001}, active galactic nuclei \cite{Sako2000}, and quasars at high redshift \cite{Simcoe2012}.

Astrophysical model calculations are required in order to infer the physical conditions of cosmic objects from their observed spectra. The photon emission from any plasma strongly depends on the ionization balances of all abundant atomic species \cite{Bryans2006,Kallman2010}. These, in turn, are governed by ionization and recombination processes. The needed cross sections and rate coefficients are obtained from theoretical calculations which bear (usually unknown) uncertainties \cite{Ballance2013a}. Thus, benchmarking by  experiments is required to test the theoretical calculations and to guide the further development of the theoretical methods.

Within our general effort to provide reliable rate coefficients for the photorecombination (PR) and electron-impact ionization (EII) of astrophysically relevant atomic ions \cite{Savin2007d,Schippers2009a,Schippers2010,Hahn2014} we here publish experimental rate coefficients for PR of Si$^{10+}$ forming Si$^{9+}$ and of Si$^{9+}$ forming Si$^{8+}$. For the measurements the electron-ion merged-beams technique was employed at a heavy-ion storage ring. In the case of Be-like Si$^{10+}$, the experimental electron-ion collision energy range was 0--2000~eV. The most significant photorecombination channels in this energy range are
{\setlength{\mathindent}{0.50cm}\begin{equation}\label{eq:Si10}
\mathrm{Si}^{10+}(2s^2)+e^- \rightarrow
\left\{\begin{array}{llll}
 \multicolumn{2}{l}{\mathrm{Si}^{9+}(2s^2\,nl) + \gamma} & &\textrm{RR}\\
 \mathrm{Si}^{9+}(2s\,2p\,nl) & \rightarrow\mathrm{Si}^{9+} + \gamma &\Delta N=0&\textrm{DR~} (2s\to2p)\\
 \mathrm{Si}^{9+}(2p^2nl)  &\rightarrow\mathrm{Si}^{9+} + \gamma &\Delta N=0&\textrm{TR~}(2s^2\to2p^2)\\
 \mathrm{Si}^{9+}(2s\,3l'\,nl) &\rightarrow\mathrm{Si}^{9+} + \gamma &\Delta N=1&\textrm{DR~} (2s\to 3l')\\
 \mathrm{Si}^{9+}(2s\,4l'\,nl) &\rightarrow\mathrm{Si}^{9+} + \gamma &\Delta N=2&\textrm{DR~} (2s\to 4l')\\
 \mathrm{Si}^{9+}(1s\,2s^2\,N'l'\,nl) &\rightarrow\mathrm{Si}^{9+} + \gamma & \Delta N>0&\textrm{DR~} (1s\to N'l')
\end{array}\right.
\end{equation}}
\noindent where $\gamma$ stands for one or more photons, and RR, DR, and TR denote radiative, dielectronic, and trielectronic recombination, respectively. The intermediate DR and TR resonance states are populated by resonant electron capture of the initially free electron into a subshell $nl$. This capture is associated with the excitation of one (in case of DR) or two (in case of TR) core electrons to a higher core level. The active core electron may be excited such that its principal quantum number $N$ does not change ($\Delta N=0$) or such that $N$ changes by 1 ($\Delta N=1$) or more ($\Delta N>1$). DR (and TR) is completed when the multiply excited intermediate level has decayed by photon emission to a bound level of the recombined ion below the first autoionization threshold.

\begin{table}[t]
\caption{\label{tab:Si10levels} Excitation energies and lifetimes of Si$^{10+}$ levels. Numbers in brackets denote powers of 10.}
\begin{indented}
    \item[]\begin{tabular}{lrrl}
 \br
       & \multicolumn{2}{c}{Excitation energy} & Lifetime \\
 Level  &  NIST \cite{Kramida2014}  & \multicolumn{2}{c}{Wang et al.\ \cite{Wang2015}} \\
        &  \multicolumn{1}{c}{(eV)} & \multicolumn{1}{c}{(eV)} & \multicolumn{1}{c}{(s)} \\
 \mr
 $1s^2\,2s^2\;^1S_0$   & 0.00000 & 0.00000 &  $\infty$  \\
 $1s^2\,2s\,2p\;^3P_0$ & 21.0528 & 21.0558 & 1.664[$+$01]$^a$ \\
 $1s^2\,2s\,2p\;^3P_1$ & 21.3431 & 21.3451 & 2.747[$-$06] \\
 $1s^2\,2s\,2p\;^3P_2$ & 21.9846 & 21.9876 & 4.729[$-$01] \\
 $1s^2\,2s\,2p\;^1P_1$ & 40.8750 & 40.8605 & 1.567[$-$10] \\
 $1s^2\,2p^2\;^3P_0$   & 55.0081 & 55.0008 & 2.034[$-$10] \\
 $1s^2\,2p^2\;^3P_1$   & 55.3582 & 55.3526 & 2.001[$-$10] \\
 $1s^2\,2p^2\;^3P_2$   & 55.9125 & 55.9056 & 1.966[$-$10] \\
 $1s^2\,2p^2\;^1D_2$   & 61.3971 & 61.3620 & 9.107[$-$10] \\
 $1s^2\,2p^2\;^1S_0$   & 75.4764 & 75.4172 & 1.042[$-$10] \\
 $1s^2\,2s\,3s\;^3S_1$ & 274.092 & 274.008 & 3.777[$-$12] \\
 $1s^2\,2s\,3s\;^1S_0$ & 277.949 & 277.884 & 1.194[$-$11] \\
 $1s^2\,2s\,3p\;^1P_1$ & 283.309 & 283.351 & 1.495[$-$12] \\
 $1s^2\,2s\,3p\;^3P_0$ & 283.330 & 283.747 & 2.894[$-$10] \\
 $1s^2\,2s\,3p\;^3P_1$ & 283.330 & 283.857 & 2.531[$-$11] \\
 $1s^2\,2s\,3p\;^3P_2$ & 283.330 & 284.005 & 2.658[$-$10] \\
 $1s^2\,2s\,3d\;^3D_1$ &         & 289.109 & 7.488[$-$13] \\
 $1s^2\,2s\,3d\;^3D_2$ & 289.137 & 289.138 & 7.515[$-$13] \\
 $1s^2\,2s\,3d\;^3D_3$ & 289.196 & 289.192 & 7.554[$-$13] \\
 $1s^2\,2s\,3d\;^1D_2$ & 292.763 & 292.731 & 1.137[$-$12] \\
 $1s\,2s^2\,2p\;^1P_1$           & 1828.618 &  &\\     	  	  	  	  	
\br
\end{tabular}
\item[]$^a$ lifetime associated with the hyperfine induced transition taken from \cite{Cheng2008a}
\end{indented}
\end{table}

For B-like Si$^{9+}$ the collision energy range was limited to 0--51~eV because of time constraints. The most important photorecombination channels in this energy range are
{\setlength{\mathindent}{0.50cm}\begin{equation}\label{eq:Si9}
\mathrm{Si}^{9+}(2s^2\,2p)+e^-\rightarrow
 \left\{\begin{array}{llll}
 \multicolumn{2}{l}{\mathrm{Si}^{8+}(2s^2\,2p\,nl) + \gamma}  & &\textrm{RR}\\
 \mathrm{Si}^{8+}(2s^2\,2p\,nl) & \rightarrow\mathrm{Si}^{8+} + \gamma &\Delta N=0&\textrm{DR~} (2p\to2p)\\
 \mathrm{Si}^{8+}(2s\,2p^2\,nl) & \rightarrow\mathrm{Si}^{8+} + \gamma &\Delta N=0&\textrm{DR~} (2s\to2p).
 \end{array}\right.
\end{equation}

Results from storage-ring experiments have been reported for many ions from the beryllium isoelectronic sequence, i.e, for
 C$^{2+}$       \cite{Fogle2005a},
 N$^{3+}$        \cite{Fogle2005a},
 O$^{4+}$        \cite{Fogle2005a},
 F$^{5+}$        \cite{Ali2013},
 Ne$^{6+}$       \cite{Orban2008a},
 Mg$^{8+}$       \cite{Schippers2004c},
 Si$^{10+}$      \cite{Orban2010},
 S$^{12+}$       \cite{Schippers2012},
 Cl$^{13+}$      \cite{Schnell2003b},
 Ti$^{18+}$     \cite{Schippers2007a,Schippers2007b},
 Fe$^{22+}$    \cite{Savin2006a},
 Ge$^{28+}$    \cite{Orlov2009}, and
 Xe$^{50+}$    \cite{Bernhardt2015a}. It is noted that the significance of TR was first discovered for Be-like Cl$^{13+}$ \cite{Schnell2003b} and subsequently confirmed for other ions from this isoelectronic sequence \cite{Schippers2004c,Orban2008a,Orban2010,Ali2013}. The boron isoelectronic sequence has been less intensely studied. DR rate coefficients from storage-ring experiments are available for
 C$^{+}$        \cite{Ali2012},
 Ne$^{3+}$      \cite{Mahmood2013},
 Mg$^{7+}$      \cite{Lestinsky2012},
 Ar$^{13+}$     \cite{Gao1995,Dewitt1996}, and
 Fe$^{21+}$     \cite{Savin2003a,Krantz2009}.
State-of-the-art theoretical DR (and TR) calculations have been carried out by Colgan \etal \cite{Colgan2003a} and by Altun \etal \cite{Altun2004a,Altun2005a}, respectively, for Be-like and B-like ions of astrophysical interest with the nuclear charge ranging up to $Z=54$. The calculated plasma rate coefficients for these ions are available from the Open ADAS database \cite{openADAS}. References to earlier theoretical work can be found in \cite{Colgan2003a,Altun2004a}.

\begin{table}[t]
\caption{\label{tab:Si9levels} Excitation energies and lifetimes of Si$^{9+}$ levels. Numbers in brackets denote powers of 10.}
\begin{indented}
    \item[]\begin{tabular}{lrrl}
 \br
       & \multicolumn{2}{c}{Excitation energy} & Lifetime \\
 Level  &  NIST \cite{Kramida2014}  & \multicolumn{2}{c}{Rynkun et al.\ \cite{Rynkun2012}} \\
        &  \multicolumn{1}{c}{(eV)} & \multicolumn{1}{c}{(eV)} & \multicolumn{1}{c}{(s)} \\
 \mr
 $1s^2\,2s^2\,2p\;^2P_{1/2}$ &             0.0000 & 0.0000  &   $\infty$    \\
 $1s^2\,2s^2\,2p\;^2P_{3/2}$ &             0.8667 & 0.8679  & 3.245[$-$01]   \\
 $1s^2\,2s\,2p^2\;^4P_{1/2}$ &            19.9627 & 19.9556 & 1.870[$-$06]   \\
 $1s^2\,2s\,2p^2\;^4P_{3/2}$ &            20.2702 & 20.2702 & 1.389[$-$05]   \\
 $1s^2\,2s\,2p^2\;^4P_{5/2}$ &            20.7128 & 20.7128 & 4.066[$-$06]   \\
 $1s^2\,2s\,2p^2\;^2D_{3/2}$ &            35.6888 & 35.7071 & 4.550[$-$10]   \\
 $1s^2\,2s\,2p^2\;^2D_{5/2}$ &            35.6926 & 35.6926 & 4.909[$-$10]   \\
 $1s^2\,2s\,2p^2\;^2S_{1/2}$ &            45.5853 & 45.6191 & 1.087[$-$10]   \\
 $1s^2\,2s\,2p^2\;^2P_{1/2}$ &            48.3588 & 48.3897 & 5.947[$-$11]   \\
 $1s^2\,2s\,2p^2\;^2P_{3/2}$ &            48.8535 & 48.8535 & 5.816[$-$11]   \\
\br
\end{tabular}
\end{indented}
\end{table}

Within the silicon isonuclear sequence of ions, experimental RR and DR results are available for Na-like Si$^{3+}$ \cite{Orban2006a,Orban2007a,Schmidt2007b}, Be-like Si$^{10+}$ \cite{Orban2010}, Li-like Si$^{11+}$ \cite{Kenntner1995,Bartsch1997} and bare Si$^{14+}$ \cite{Gao1997}. In addition, detailed measurements of recombination resonances have been conducted using x-ray spectroscopy of silicon ions in an electron-beam ion-trap (EBIT) \cite{Baumann2014}. The previous experimental data for Si$^{10+}$ \cite{Orban2010} cover a limited range of electron-ion collision energies of 0--43~eV. Thus, these data comprise only $\Delta N=0$ DR ($2s\to2p$) and part of the $\Delta N=0$ TR ($2s^2\to2p^2$). In contrast, the present Si$^{10+}$ measurements were carried out over a much wider collision energy range extending up to 2000~eV. This energy range covers all DR resonances including those associated with K-shell excitation. Our resulting Si$^{10+}$ plasma rate coefficient is accordingly valid for much higher temperatures including those where Si$^{10+}$ forms in a collisionally ionized plasma. It should be noted, however, that our Si$^{9+}$ DR plasma rate coefficient is only valid for lower temperatures including those where Si$^{9+}$ forms in a photoionized plasma.

\section{Experiment}\label{sec:exp}

The PR measurements were performed at the heavy-ion storage-ring TSR \cite{Grieser2012} of the Max-Planck-Institut f\"ur Kernphysik (MPIK) in Heidelberg, Germany.
The experimental procedures have been described before \cite{Kilgus1992,Lampert1996,Schippers2001c,Novotny2012}. Therefore, we here focus on the details that are specific to the present measurements.

$^{29}$Si$^{10+}$ and $^{28}$Si$^{9+}$ were provided by the MPIK tandem and linear accelerators and injected into TSR at energies of about 120~ and 100~MeV,
respectively. The choice of the less abundant isotope $^{29}$Si with a natural abundance of only 4.7\% for the measurements with Si$^{10+}$ is motivated below (\sref{sec:meta}). The stored ion currents were typically 10--20~$\mu$A for Si$^{10+}$ and 1~$\mu$A for Si$^{9+}$ after injection and from then on decreased exponentially as a function of storage time. The TSR electron cooler was used for phase-space cooling of the ion beam and as an electron target for the recombination measurements. The high-resolution electron target, which is installed at TSR in addition to the cooler, was not used for the present measurements since its photocathode provides much lower electron currents and consequently much lower recombination signal rates than the thermionic cathode of the cooler. The first dipole magnet behind the electron cooler was used to separate the recombined ions from the circulating beam. The recombined ions were detected by an appropriately placed single-particle detector \cite{Rinn1982} with nearly 100\% efficiency.

During electron cooling the electron energy was set to the cooling energy $E_\mathrm{cool}$ where electrons and ions move with the same velocity. This corresponds to zero collision energy in the electron-ion center-of-mass frame. The cooling energy was 2076~eV for Si$^{10+}$ and 1788~eV for Si$^{9+}$. After injection, the ions were  continuously cooled for 2~s to reach optimal experimental conditions. For the recombination measurements the electron-ion collision energy $E$ was detuned from zero by changing the cooler cathode voltage. Each energy scan comprised typically $k_\mathrm{max}=300$ preselected electron-ion collision energies $E_\mathrm{meas}^{(k)}$ with $k=1,\ldots,k_\mathrm{max}$. Between any two measurement steps the electron-ion collision energy was set to a reference value $E_\mathrm{ref}$  and then to $E_\mathrm{cool}$ to determine background and to maintain the ion beam quality, respectively. The intermediate cooling step was only applied at collision energies below 270~eV. It was left out at higher energies since the associated voltage jumps became too large for maintaining well controlled experimental conditions. There was a  waiting time interval of 17~ms duration after each change of $E$ to allow the power supplies to settle to their new voltages. Then data were taken at $E_\mathrm{meas}^{(k)}$, $E_\mathrm{ref}$, and $E_\mathrm{cool}$ for a dwell time interval of 25--50~ms duration. The recombination signal recorded at $E_\mathrm{ref}$ was used for background determination. The entire injection-cooling-scanning sequence was repeated hundreds to thousands of times until the statistical uncertainties were reduced to the desired level. Multiple collision-energy scans with typically 50\% overlap were used to cover the electron-ion collision energy ranges 0--2000 eV for Si$^{10+}$ and 0--50~eV for Si$^{9+}$.

Merged-beams rate coefficients (see e.g.~\cite{Mueller2015b}) were derived by normalizing the measured recombination count rate (after subtraction of background due to electron capture by the stored ions from residual gas particles) to the electron density in the cooler and to the stored-ion current and by taking the geometrical overlap of both beams into account \cite{Schippers2001c}. In addition, a correction procedure was applied that accounts for the nonzero angles between the electron and the ion beam in the toroidal sections of the electron cooler \cite{Lampert1996}. The ion current was measured with a DC beam transformer with an accuracy of 1--2~$\mu$A. Additionally, in order to measure also Si$^{10+}$ currents below 1~$\mu$A, the count rate from the TSR beam profile monitor (BPM) was used as a proxy for the ion current. The BPM is based on residual gas ionization \cite{Hochadel1994a} and there is a linear relationship between BPM count rate and stored ion current. The constant of proportionality was calibrated repeatedly against the current transformer measurement at high stored ion currents. For Si$^{9+}$, however, the ion currents were too low for a reliable calibration of the BPM signal. Therefore, the Si$^{9+}$ rate coefficient scale was normalized to the absolute recombination rate coefficient at zero electron-ion collision energy. The latter was determined from a separate measurement of the storage lifetime of the ion beam following the procedures of \cite{Novotny2012}, which were also employed in recent recombination experiments with W$^{20+}$ \cite{Schippers2011} and W$^{18+}$ \cite{Spruck2014a} ions, where only very low ion currents were available. Here we tested the reliability of this method by performing a cross check with Si$^{10+}$ ions. We found that the Si$^{10+}$ recombination rate-coefficients from the beam-lifetime and the ion-current measurements agreed with each other to within the experimental uncertainties.
The uncertainty of the Si$^{9+}$ merged-beams rate-coefficient scale is estimated to amount to $\pm 25\%$ (at 90\% confidence level) \cite{Lampert1996}. In case of the Be-like Si$^{10+}$ primary ion there is an additional $7\%$ uncertainty due to an unknown fraction of metastable ions in the primary beam as discussed below. Adding the uncertainties in quadrature results in a $\pm26\%$ uncertainty of the Si$^{10+}$ merged-beams rate-coefficient scale. The systematic uncertainty of the electron-ion collision energy scale is negligible at electron-ion collision energies below $\sim$1~eV and increases with increasing energy. A conservative estimate \cite{Kilgus1992} yields systematic uncertainties of 0.4, 0.7, and 1.5 eV at electron-ion collision energies of 40, 200, and  1700 eV, respectively.

\subsection{Metastable primary ions} \label{sec:meta}

Be-like ions are known to have long-lived $2s2p~^3P_J$ levels ($J=0,1,2$) which might have been present in the ion beam used for the measurements described here. The predicted lifetime of the $^{28}$Si$^{10+}(2s2p)~^3P_0$, $^3P_1$ and $^3P_2$ levels are several 100 days \cite{Fritzsche2015}, 2.747~$\mu$s and 0.4729~s \cite{Wang2015} due to the dominant two-photon (E1M1), intercombination (E1), and magnetic quadrupole (M2) transitions, respectively. For nuclei with a nonzero magnetic moment, hyperfine quenching shortens the $^3P_0$ lifetime by 6 orders of magnitude. Recent calculations yield a value of about 17~s for the hyperfine induced (HFI) lifetime of the  $^{29}$Si$^{10+}$($2s\,2p\;^3P_0$) level \cite{Cheng2008a,Andersson2009}.

\begin{figure}[b]
 \begin{indented}
 \item[]\includegraphics[width=0.5\textwidth]{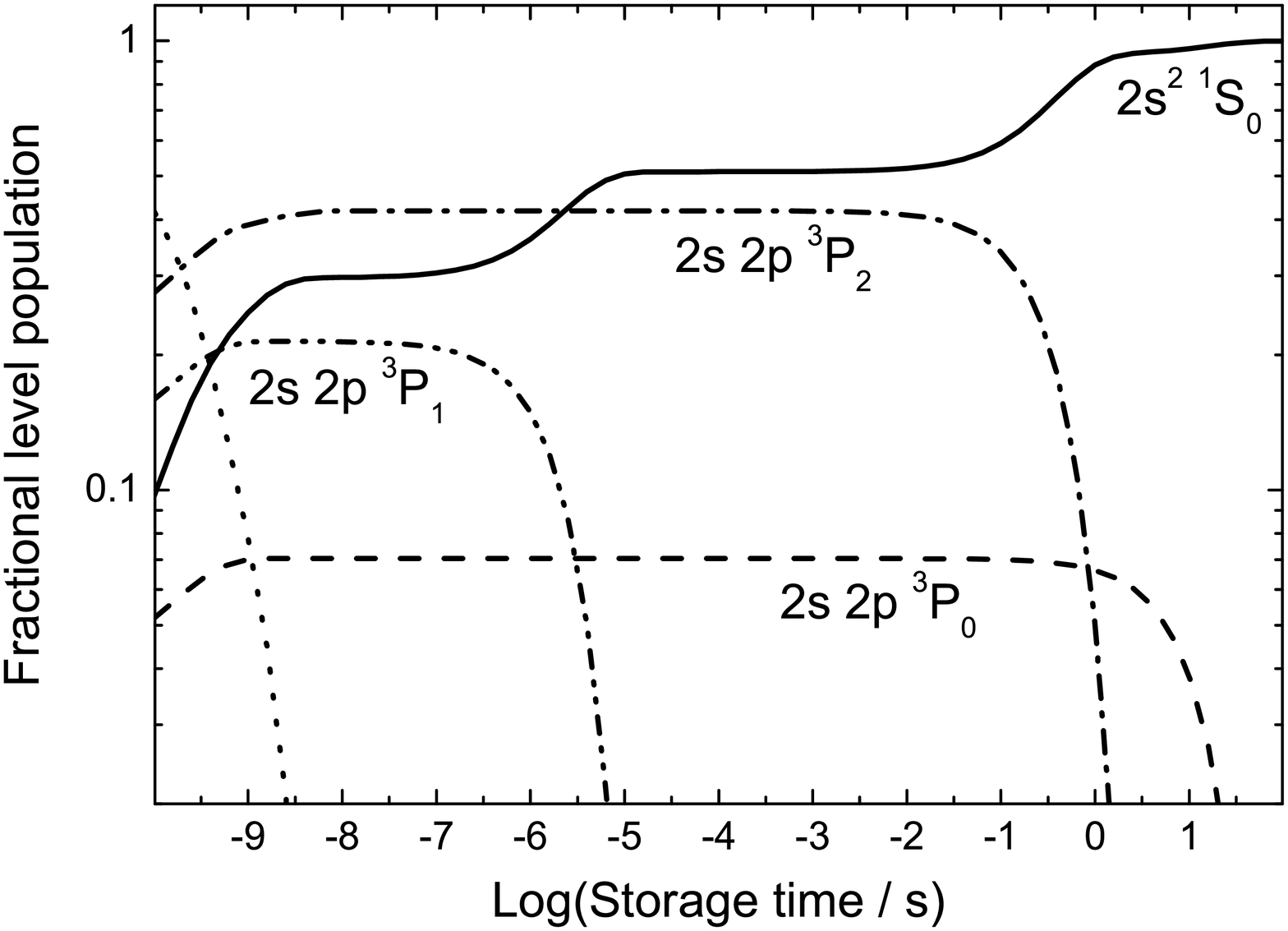}
 \end{indented}
\caption{\label{fig:Si10pop}Calculated fractional $^{29}$Si$^{10+}$ level populations as a function of storage time using theoretical transition rates \cite{Wang2015} (see also \tref{tab:Si10levels}) for the 116 lowest Si$^{10+}$ levels. In addition, the rate for the hyperfine induced $2s\,2p\;^3P_0 \to 2s^2\;^1S_0$ transition ($A_\mathrm{HFI} = 0.06011$~s$^{-1}$) was taken from \cite{Cheng2008a}. The different lines correspond to the following levels: $2s^2\;^1S_0$ (\full), $2s\,2p\;^3P_0$ (\broken), $2s\,2p\;^3P_1$ (\dashddot), and $2s\,2p\;^3P_2$ (\chain). The dotted curve (\dotted) is the sum of the fractional populations of the remaining 112 levels. }
\end{figure}

In fact, one motivation for the present recombination measurement was to derive the HFI $^3P_0$ lifetime using the same approach as already successfully applied for the HFI  lifetime of the $2s\,2p\;^3P_0$ level in Be-like $^{47}$Ti$^{18+}$ \cite{Schippers2007a} and $^{33}$S$^{12+}$ \cite{Schippers2012}. This approach requires the identification of DR resonances of $^3P_0$ parent ions by comparing recombination spectra with different amounts of the Si$^{10+}$($2s\,2p\;^3P_0$) parent ion. If such a resonance is present the visible resonance strength in the measured $^{29}$Si$^{10+}$ spectrum would be suppressed with increasing storage time due to the decaying $^3P_0$ level. However, no significantly strong DR resonances associated with the initially metastable $^3P_0$ level were found. This is attributed to the fact that DR resonances of the $^3P_0$ parent ions have been found to be weak and to be blended with the rich DR resonance structure of the $^1S_0$ ground state parent ions.

We estimated the $^3P_0$ fraction in the parent beam by solving a set of rate equations \cite{Lestinsky2012} that accounts for the population and depopulation of the various $^{29}$Si$^{10+}$ levels by radiative transitions. The corresponding rates were taken from \cite{Wang2015,Cheng2008a} (\tref{tab:Si10levels}), and a Boltzmann distribution with a temperature of about 2100 eV (corresponding to the ion energy at the last stripping foil in the accelerator) was chosen as the initial population of the levels. The resulting level populations are shown in \fref{fig:Si10pop}. They turned out to be rather insensitive to the choice of initial populations. After an ion-storage time of 2~s, only two levels remain significantly populated, i.e., the $2s^2\;^1S_0$ ground level and the $2s\,2p\;^3P_0$ first excited level with their fractional populations amounting to 93\% and 7\%, respectively. At longer times, the $2s\,2p\;^3P_0$ level is depopulated by the $2s\,2p\;^3P_0\to 2s^2\;^1S_0$ HFI decay in $^{29}$Si$^{10+}$. Since data collection in the experiment started only after an initial cooling time of 2~s, we assume in the following that the fraction of ions in the $2s\,2p\;^3P_0$ level was at most 7\%. This is a conservative estimate since the $^3P_0$ fraction is predicted to have dropped to below 1\% during the time of at least 30~s (see above) that was required to complete an energy scan.

The longest lived excited level of Si$^{9+}$ is the $2s^2\,2p\;^2P_{3/2}$ level with a calculated lifetime of 0.324~s \cite{Rynkun2012} (\tref{tab:Si9levels}). This lifetime is much shorter than the initial beam-cooling time of 2~s. It is thus assumed that all Si$^{9+}$ ions were in their ground level when data taking was started.

\section{Results}

\subsection{Merged-beams recombination rate coefficient for Si$^{10+}$}

\begin{figure}[t]
 \begin{indented}
 \item[]\includegraphics[width=\figurewidth]{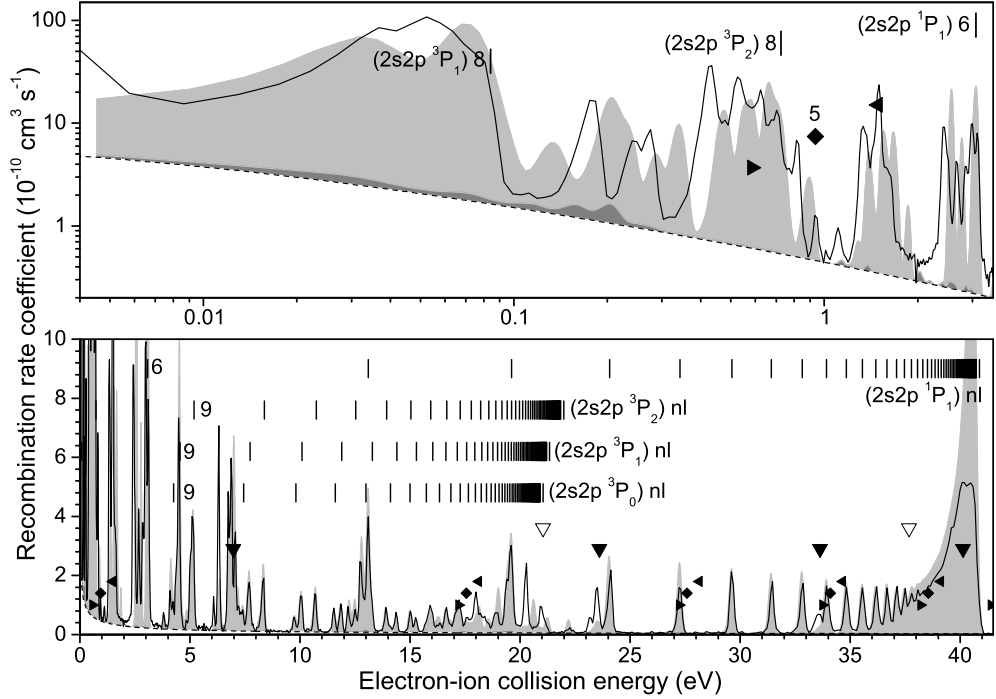}
 \end{indented}
\caption{\label{fig:Si10DN0}Measured merged-beams rate coefficient (solid black line) for PR of Si$^{10+}$ in the energy range of the $2 \to 2$ $\Delta N=0$ DR resonances. The vertical bars denote  DR resonance positions  for ground-level ions calculated using equation \eref{eq:DR} with $q=10$ and with the excitation energies taken from \tref{tab:Si10levels} (NIST values). Calculated energies of TR resonances associated with $2s^2 +e^-\to 2p^2~^3P_{0,1,2}$, $^1D_2$ and $^1S_0$ excitations are marked by the following symbols: $\blacktriangleright, \vardiamond, \blacktriangleleft$, $\blacktriangledown$ and $\triangledown$, respectively. The calculated RR rate coefficient is shown as a dashed line. Results from DR and TR calculations with the \textsc{Autostructure} code are shown as shaded areas. In these calculations an ion beam composition of 93\% $2s^2\;^1S_0$ ground level and 7\% $2s\,2p\;^3P_0$ metastable level was assumed. DR resonances associated with the excitation of the $2s^2\;^1S_0$ ground level (light shaded) and the $2s\,2p\;^3P_0$ metastable level are shown separately (dark shaded, only visible in the top panel).}
\end{figure}

The measured merged-beams rate coefficients for the recombination of Si$^{10+}$ ions are displayed in Figures \ref{fig:Si10DN0}, \ref{fig:Si10DN1}, and \ref{fig:Si10KLX} which comprise the energy ranges of DR associated with $2s\to2p$ ($\Delta N=0$), $2s\to 3l,4l$ ($\Delta N=1,2$), and $1s\to N'l'$ core excitations, respectively. The strongest resonances are associated with the $2s^2\;^1S_0\to 2s\,2p\;^1P_1$ core excitation with an excitation energy $E_\mathrm{exci} = 40.875$~eV. In general, recombination resonance positions can be estimated from the Rydberg formula
\begin{equation}\label{eq:DR}
    E_\mathrm{res}(n) = E_\mathrm{exci}-\mathcal{R}\frac{q^2}{n^2}
\end{equation}
with $\mathcal{R}  \approx 13.6057$~eV, the primary ion charge state $q=10$ and the excitation energies $E_\mathrm{exci}$ from \tref{tab:Si10levels}. The resulting calculated resonance positions are indicated by vertical bars in figures \ref{fig:Si10DN0} and \ref{fig:Si10DN1}. Obviously the simple Rydberg formula works reasonably well for high-$n$ resonances where the interaction between the Rydberg electron and the excited core is only weak. The low-$n$ resonance structure, however, is dominated by fine-structure effects and the assignment of the measured structure becomes more complicated. More detailed insight is provided by comparing the measured data with results of theoretical calculations. To this end we have employed the \textsc{Autostructure} code \cite{Badnell2011}.  Our theoretical results for $\Delta N=0$ DR and TR are shown in \fref{fig:Si10DN0} as shaded curves. For the comparison the theoretical cross sections were convoluted with the experimental electron velocity distribution which is characterized by the temperatures $T_\parallel$ and $T_\perp$ parallel and perpendicular to the propagation direction of the electron beam \cite{Kilgus1992}.  For the comparison in \fref{fig:Si10DN0}, $k_\mathrm{B}T_\parallel = 93$~$\mu$eV and $k_\mathrm{B}T_\perp = 11$~meV were used (with $k_\mathrm{B}$ being Boltzmann's constant) as determined by fits of theoretical line shapes to narrow DR resonances at electron-ion collision energies of 0.17 and 6.4 eV.

The calculated and measured merged-beams rate coefficients agree very well with one another over almost the entire energy range displayed in \fref{fig:Si10DN0}. At energies below $\sim$3~eV, there are small deviations in resonance positions and strengths. In addition,  close to the $2s\,2p\,(^1P_1)\,nl$ series limit, the calculated rate coefficient is much larger than the experimental result. The origin of this latter discrepancy is well understood. It is due to field ionization of weakly bound high-$n$ Rydberg states which occurs primarily in the dipole bending magnets of the storage ring \cite{Schippers2001c}. This field ionization effect has to be accounted for in the derivation of the plasma rate coefficient.

Apart from the dominating $2s\,2p\,(^1P_1)\,nl$ Rydberg series, there are the weaker $2s\,2p\,(^3P)\,nl$ series with their series limits at around 21.5~eV (\tref{tab:Si10levels}) and a few isolated TR resonances such as the $2p^2\,(^1S_0)\,5l$ resonance at 21.0 eV, the $2p^2\,(^1D_2)\,6l$ resonance at 23.6 eV, and the $2p^2\,(^1D_2)\,7l$ resonance at 33.6 eV. The role of TR in low-energy recombination of Si$^{10+}$ has been studied already thoroughly by Orban et al.~\cite{Orban2010}. They found that a large part of the resonance structure below 9~eV and in the range 15--25~eV can be attributed to TR and, thus, confirmed earlier findings on the importance of TR for Be-like ions \cite{Schnell2003b}.

The present \textsc{Autostructure} calculations were carried out for primary Si$^{10+}$ ions in the $2s^2\;^1S_0$ ground level (light shaded area in \fref{fig:Si10DN0}) and in the long-lived $2s\,2p\;^3P_0$ metastable level (dark shaded area in \fref{fig:Si10DN0}) assuming a ratio of 93:7 for the two beam components (cf.\ section~\ref{sec:meta}). The calculations suggest that the contribution of the initial metastable ions to the measured $\Delta N=0$ DR+TR rate coefficient is practically negligible. Nevertheless, in the derivation of a pure ground-level plasma rate coefficient the (assumed) 7\% metastable fraction introduces a corresponding uncertainty in the normalization to the ion current which has been taken into account in our error budget as described above.

\begin{figure}[t]
 \begin{indented}
 \item[] \includegraphics[width=\figurewidth]{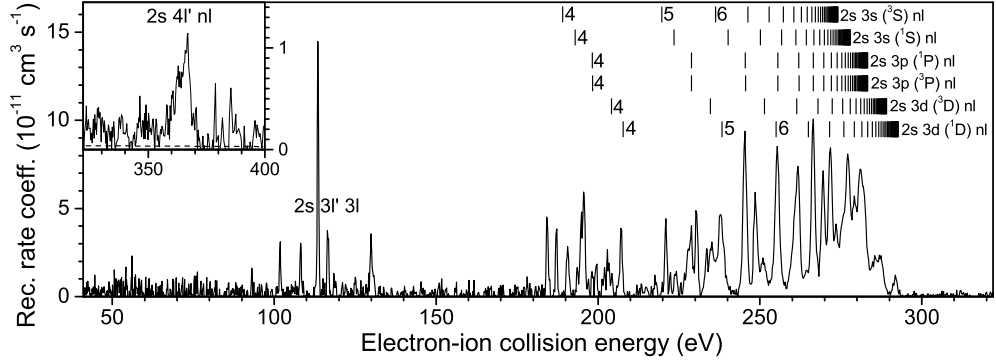}
  \end{indented}
\caption{\label{fig:Si10DN1} Measured merged-beams rate coefficient (solid black line) for PR of Si$^{10+}$ in the energy range of DR resonances associated with $2 \to 3$ (main panel) and $2 \to 4$ (inset) core excitations. The vertical bars denote  DR resonance positions of ground-level ions calculated using equation \eref{eq:DR} with $q=10$ and the excitation energies from \tref{tab:Si10levels} (NIST values).}
\end{figure}

In contrast to the $\Delta N=0$ resonances discussed so far, the DR resonance structure associated with $2s\to3l$ $\Delta N=1$ core excitations (\fref{fig:Si10DN1}) is less regular. This is due to the many mutually overlapping Rydberg series of DR resonances. The $\Delta N=1$ series limits are between 274 and 293~eV (\tref{tab:Si10levels}). At higher energies of about 370 eV, resonances associated with $2s\to4l$ $\Delta N=2$ core excitations are visible. They are an order of magnitude smaller than the $\Delta N=1$ resonances. Contributions associated with even higher excitations were not observed. They can be expected to be negligibly small. The cumulative strength of the high-$n$ $\Delta N=1$ resonances does not pile up as much as for the $\Delta N=0$ Rydberg series because the $\Delta N=1$ resonance strength decreases much faster with increasing $n$ than the $\Delta N=0$ resonance strength. Therefore, $\Delta N=1$ DR is much less affected by field ionization than $\Delta N=0$ DR.

\begin{figure}[t]
\begin{indented}
 \item[]  \includegraphics[width=\figurewidth]{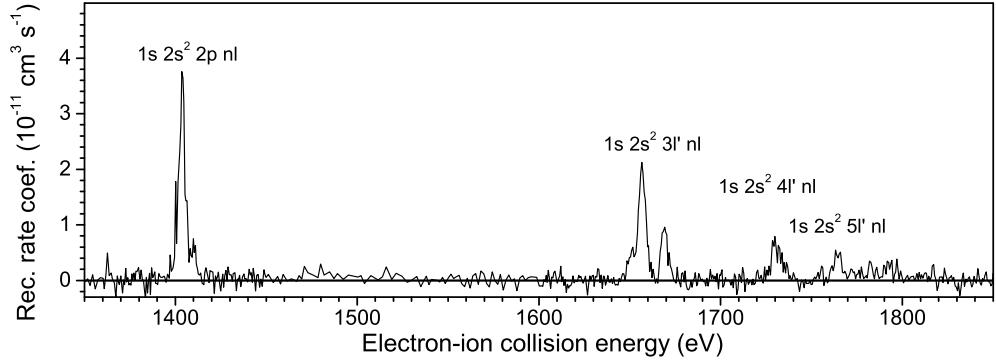}
  \end{indented}
\caption{\label{fig:Si10KLX} Measured merged-beams rate coefficient (solid black line) for PR of Si$^{10+}$ in the energy range of DR resonances associated with K-shell excitations. Resonance groups are labelled by the according $1s^2\,2s^2 \to 1s\,2s^2\,N'l'$ core excitations.}
\end{figure}

The highest electron-ion collision energies studied were in the range 1300--2000~eV where DR resonances occur that are associated with the excitation of a $1s$ electron to a higher principal shell $N'$ (\fref{fig:Si10KLX}). The $1s\,2s^2\,N'l'\,nl$ DR resonance strengths decrease rapidly with increasing $N'$ and increasing $n$. There is no significant DR resonance strength beyond an electron-ion collision energy of $\sim$1800~eV since DR cannot occur at energies beyond the threshold for direct K-shell ionization of the Si$^{10+}$ parent ion.

\subsection{Merged-beams recombination rate coefficient for Si$^{9+}$}

Experimental and theoretical data for RR and $\Delta N=0$ DR of boronlike Si$^{9+}$ ions are presented in \fref{fig:Si9DN0}. The theoretical cross sections were calculated using the \textsc{Autostructure} code and merged-beams rate coefficients were subsequently derived by convolution with the experimental electron velocity distribution using $k_\mathrm{B}T_\parallel = 180$~$\mu$eV and $k_\mathrm{B}T_\perp = 12$~meV. The Si$^{9+}$ DR spectrum exhibits three strong Rydberg series associated with the formation of doublet terms of $2s\,2p^2$ core excited levels (\tref{tab:Si9levels}).  At energies below 1 eV DR Rydberg resonances can be formed via $2s^2\,2p\;^2P_{1/2} \to 2s^2\,2p\;^2P_{3/2}$ fine-structure core excitations \cite{Savin1997} with the corresponding series limit occurring at 0.867~eV. As for Si$^{10+}$ the theoretical calculations for Si$^{9+}$ agree also well the experimental data, except for the resonance structure below $\sim$2~eV which is notoriously difficult to calculate accurately because the multiply excited levels just above the autoionization threshold of the recombined ion are subject to strong correlation effects. The differences close to the Rydberg series limits are due to field ionization effects as already discussed above for Si$^{10+}$.

\begin{figure}[t]
  \begin{indented}
 \item[] \includegraphics[width=\figurewidth]{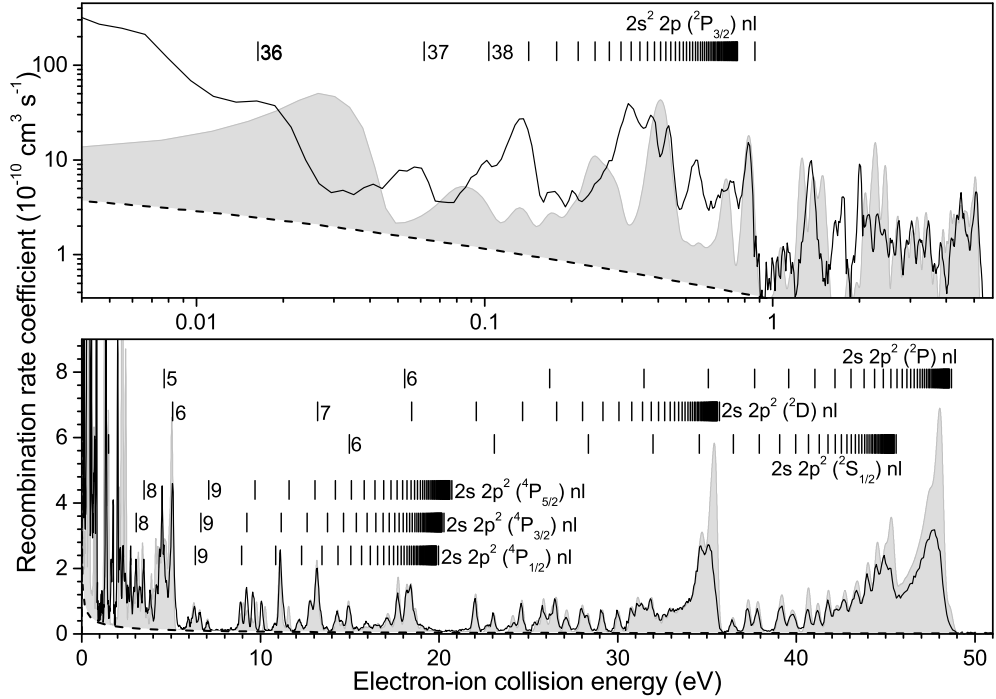}
  \end{indented}
\caption{\label{fig:Si9DN0} Measured merged-beams rate coefficient (solid black line) for PR of Si$^{9+}$ in the energy range of DR resonances associated with intra L-shell $\Delta N=0$ core excitations. The black vertical bars denote DR resonance positions calculated using equation \eref{eq:DR} with $q=9$ and with the excitation energies taken from \tref{tab:Si9levels} (NIST values). Results from DR calculations with the \textsc{Autostructure} code are shown as a shaded curve on top of the small calculated RR-rate coefficient (dashed line).}
\end{figure}

\subsection{Plasma rate coefficients}

\begin{figure}[t]
   \begin{indented}
 \item[] \includegraphics[width=0.5\textwidth]{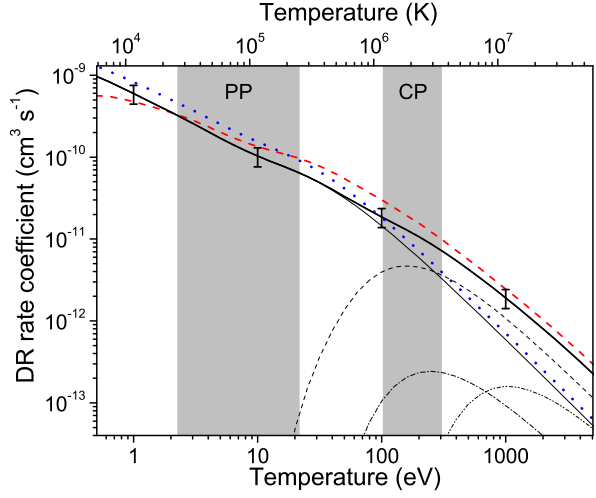}
  \end{indented}
\caption{\label{fig:Si10PRC}(Color online) Rate coefficients for DR of Si$^{10+}$ ($2s^2~^1S_0$) ions in a plasma. The thick full line is our present experimentally derived rate coefficient. The error bars denote the $\pm 26\%$ systematic uncertainty at a 90\% confidence level. DR contributions associated with $2\to2$, $2\to3$, $2 \to 4$ and $1\to N'$ excitations are given as straight, dashed , dash-dotted  and dot-dash-dotted thin lines. The (blue) dotted line is the result for $\Delta N=0$ DR by Orban \etal \cite{Orban2010}. The most recent theoretical result \cite{Colgan2003a} is shown as (red) thick dashed line. Approximate temperature ranges where Si$^{10+}$ is expected to form in photoionized plasmas (PP) \cite{Kallman2001} and collisionally ionized plasmas (CP) \cite{Bryans2006} are indicated as gray shaded areas.}
\end{figure}

\begin{figure}[t]
   \begin{indented}
 \item[] \includegraphics[width=0.5\textwidth]{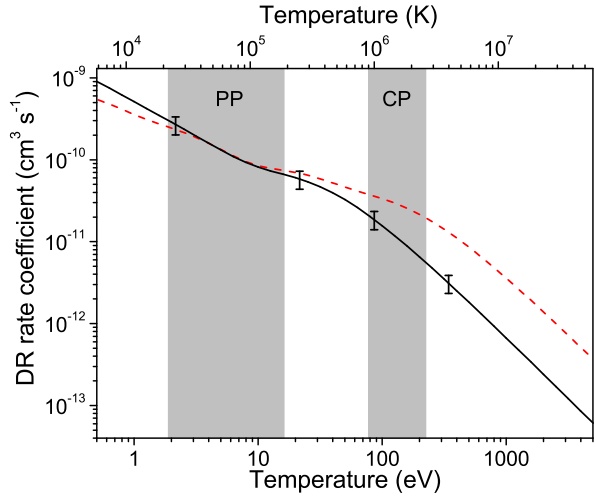}
  \end{indented}
\caption{\label{fig:Si9PRC}(Color online) Rate coefficients for DR of Si$^{9+}$($2s^2~2p^2P_{1/2}$) ions in a plasma. The thick full line is our present experimentally derived rate $\Delta N=0$ DR coefficient. The error bars denote the $\pm 25\%$ uncertainty at a 90\% confidence level. The most recent theoretical results \cite{Altun2004a} comprising $\Delta N=0$ and $\Delta N>0$-DR is shown as a dashed line. Approximate temperature ranges where Si$^{9+}$ is expected to form in photoionized plasmas (PP) \cite{Kallman2001} and collisionally ionized plasmas (CP) \cite{Bryans2006} are indicated as gray shaded areas.}
\end{figure}

For the derivation of plasma rate coefficients of Si$^{9+}$ and Si$^{10+}$ ions from the measured merged-beams recombination rate coefficients we follow the procedure described by Schippers \etal \cite{Schippers2001c}. Accordingly, field ionization is accounted for by replacing the measured rate-coefficients by appropriately scaled \textsc{Autostructure} results in the energy ranges close to the $\Delta N=0$ Rydberg series limits. No such corrections are required for the $\Delta N>0$ series limits, which are much less affected by field ionization as explained above. Plasma rate coefficients for DR(+TR) of ground-level ions were then derived by first subtracting the calculated RR rate coefficients from the measured data and by then convoluting the remaining DR(+TR) merged-beams rate coefficients by an isotropic Maxwellian. The resulting Si$^{10+}$ and Si$^{9+}$ DR rate coefficients in a plasma are plotted in figures \ref{fig:Si10PRC} and \ref{fig:Si9PRC} as functions of plasma temperature. Their systematic uncertainties of 26\% and 25\%, respectively, at a 90\% confidence level, correspond directly to the systematic uncertainties of the underlying experimental merged-beams rate coefficients.

The present experimentally derived Si$ ^{10+}$ plasma DR+TR rate coefficient agrees with the most recent theoretical result \cite{Colgan2003a} within the experimental uncertainties over almost the entire plotted (\fref{fig:Si10PRC}) temperature range of 0.5--5000 eV. In particular, there is good agreement in the temperature ranges where Si$^{10+}$ forms in photoionized plasmas (PP) and collisionally ionized plasmas (CP). These temperature ranges are shaded in \fref{fig:Si10PRC} and have been estimated on the basis of charge balance calculations by Kallman and Bautista \cite{Kallman2001} and by Bryans \etal \cite{Bryans2006}, respectively.  Also shown in \fref{fig:Si10PRC} are the individual contributions by the processes listed in equation \ref{eq:Si10}. Clearly, $\Delta N>0$ DR contributes significantly to the total DR rate coefficient in the CP temperature range.  The previous experimentally derived result of Orban \etal \cite{Orban2010} for $\Delta N=0$ DR+TR (dotted line in \fref{fig:Si10PRC}) agrees with the present $\Delta N=0$ DR+TR rate coefficient within the combined experimental uncertainties.

Within the 25\% experimental uncertainty (at a 90\% confidence level), the Si$^{9+}$ DR rate coefficient agrees with the most recent theoretical result \cite{Altun2004a} only for temperatures between 1.5 and 30~eV comprising the entire PP temperature range (\fref{fig:Si9PRC}). At higher temperatures the experimental result is significantly lower than the theoretical prediction. This is due to the neglect of $\Delta N>0$ DR in the experimentally derived rate coefficient of Si$^{9+}$. Because of time constraints there were no measurements in the corresponding electron-ion collision energy ranges.

For convenient use of our results in plasma modeling codes, we provide a simple parametrization where we have fitted the function
\begin{equation} \label{eq:PRCfit}
\alpha_\mathrm{DR}(T) = T^{-3/2} \sum_{i} c_i \exp\left (-E_i /T \right )
\end{equation}
to our experimentally derived DR(+TR) plasma rate coefficients. The fit parameters $c_i$ and $E_i$ are listed in \tref{tab:PRCfit}.
For plasma temperatures $T$ between $10^2$ and $5\times10^7$ K, the fits deviate by less than 2\% from the experimentally derived curves.

 \begin{table}[t]
\caption{\label{tab:PRCfit} Parameters for the parametrization (equation \ref{eq:PRCfit}) of the experimentally derived rate coefficients for DR+TR of Si$^{10+}$($2s^2\;^1S_0$) and for $\Delta N=0$ DR of Si$^{9+}$($2s^2\,2p\;^2P_{1/2}$) in a plasma. Numbers in square brackets denote powers of 10. The parameters are valid for plasma temperatures between $10^2$ and $5\times 10^7$~K.}
\begin{indented}
    \item[]\begin{tabular}{rrrcrr}
 \br
    & \multicolumn{2}{c}{Si$^{10+}$} &~~~&\multicolumn{2}{c}{Si$^{9+}$} \\
$i$ & $c_i$ (cm$^3$~s$^{-1}$~K$^{3/2}$) & $E_i$ (K) & &$c_i$ (cm$^3$~s$^{-1}$~K$^{3/2}$) & $E_i$ (K) \\
\mr
 1&	1.932[-5]&	2.996[1] & &	4.438[-5]&	3.837[1]\\
 2&	4.999[-6]&	5.778[1] & &	4.577[-5]&	9.834[1]\\
 3&	1.132[-4]&	5.823[2] & &	5.189[-5]&	9.261[2]\\
 4&	9.121[-5]&	1.477[3] & &	5.511[-4]&	4.656[3]\\
 5&	7.094[-4]&	6.760[3] & &	1.183[-3]&	2.507[4]\\
 6&	1.576[-3]&	2.818[4] & &	3.148[-3]&	1.161[5]\\
 7&	3.738[-3]&	1.106[5] & &	2.219[-2]&	4.584[5]\\
 8&	1.801[-2]&	4.200[5] & & &\\
 9&	5.727[-2]&	2.864[6] & & & \\
10&	3.160[-2]&	1.701[7] & & &\\
\br
\end{tabular}
\end{indented}
\end{table}

\section{Summary and conclusions}

Rate coefficients for dielectronic recombination of Be-like Si$^{10+}$ and B-like Si$^{9+}$ have been measured at a heavy ion storage ring. For Si$^{10+}$ the energy range has been greatly extended compared to the previous measurements by Orban \etal \cite{Orban2010}. The experimental energy range covered by the present experiment with Si$^{10+}$ comprises the highest-energy DR resonances associated with K-shell core excitation. Correspondingly, the experimentally derived plasma rate coefficient for DR+TR of Si$^{10+}$ is valid for all temperatures. Within the experimental uncertainties it agrees with the most recent theoretical result \cite{Colgan2003a}. The present result for $\Delta N=0$ DR+TR also agrees with the previous experimentally derived rate coefficient \cite{Orban2010} which was measured at another heavy-ion storage ring. This agreement (within the combined experimental uncertainties) demonstrates the reliability of the storage-ring technique. One of the hallmarks of this technique is that it allows for the preparation of ions in well defined energy levels. This has been exploited in particular for reducing the Si$^{10+}$ ion beam contamination by long-lived $2s\,2p\;^3P$ metastable levels to almost insignificance.

For Si$^{9+}$ only low-energy $\Delta N=0$ DR was measured. The experimentally derived plasma rate coefficient agrees with the most recent theoretical result \cite{Altun2004a} in the temperature range where Si$^{9+}$ forms in a photoionized plasma. Our result may thus be applied in charge balance calculations for, e.g., active galactic nuclei \cite{Mueller2012}.

Our experimental results benchmark state-of-the-art theoretical calculations for few-electron ions. The good agreement found provides confidence in the theoretical methods typically used for generation of DR(TR) plasma rate coefficients. It should be kept in mind however, that the accuracy of these theoretical methods is much more limited for many-electron ions as is shown by recent investigations of electron-ion recombination of complex open-$4f$-shell ions \cite{Mueller2015b,Schippers2011,Badnell2012,Dzuba2013,Spruck2014a}.

\ack

We would like to thank the MPIK accelerator and TSR crews for their excellent support during the experiments. Financial support by the Max-Planck-Gesellschaft and by the Deutsche Forschungsgemeinschaft (DFG, contract no.\ Schi\,378/8-1) is gratefully acknowledged. MH, ON, and DWS were supported in part by the NASA Solar Heliospheric Physics program.

\section*{References}


\providecommand{\newblock}{}

\end{document}